# The Empirical Network Analysis of Website Visitors

Mohammed K.A Kaabar

IEEE Student Member, IEEE Computer Society Member , IEEE Women Engineering Society Member and IEEE Comunications Society Member

Social Network Analysis by Dr.Lada Adamic
University of Michigan – Coursera – Signature Track
mohdkaabar@gmail.com

*Abstract*— **This paper explores how to analyze empirically a network of website visitors from several countries in the world. While exploring this huge network of website visitors worldwide, this paper shows an empirical data analysis with a visualization of how data has been analyzed and interpreted. By evaluating the methods used in analyzing and interpreting these data, this paper provides the required knowledge to empirically analyze a set of various obtained data from website visitors with different browsers and IP-addresses.**

## I. INTRODUCTION

Nowadays, there is a significant need for analyzing a network of website visitors by using the obtained data from the visitors of website including sub-data such as returning visits, visit length, page loads, IP-addresses, browsers, operating systems, countries with different regions and cities. Therefore, applying the methods used in social network analysis (SNA) for analyzing and interpreting data of website visitors, can give us insight and overview of how SNA can be used beneficially in several applications of web development. In this social network analysis project, three main steps will be followed in order to be able to satisfy the requirements of the empirical network analysis of website visitors. First of all, in section II, the obtained data from website will be discussed and documented with the applied source code as well as the number of nodes and edges will be determined using these data. Then, in section III, the data analysis of website visitors will be explained in more details using various methods from SNA course and some methods outside the SNA course will be also discussed and how can be used to get a good visualization of the empirical network of website visitors worldwide. After that, in section IV, these obtained data from website visitors will be interpreted, and the limitations of these data will be also addressed in order to see how the new insights can be gained from the empirical network analysis of website visitors. Ultimately, this SNA project is depended on an educational website founded by Mohammed Kaabar in September, 2012 where this website contains eight different sections which gives an overview about the personal information of Mohammed Kaabar, attended and participated conferences, memberships, forum, articles and projects in some interesting topics of electrical engineering such as antennas & wave propagation, modern electronics, pattern recognition and information about some other courses in engineering in general and electrical engineering in particular. Fig.1 below shows a picture of Mohammed Kaabar Website.

## II. OBTAINING DATA

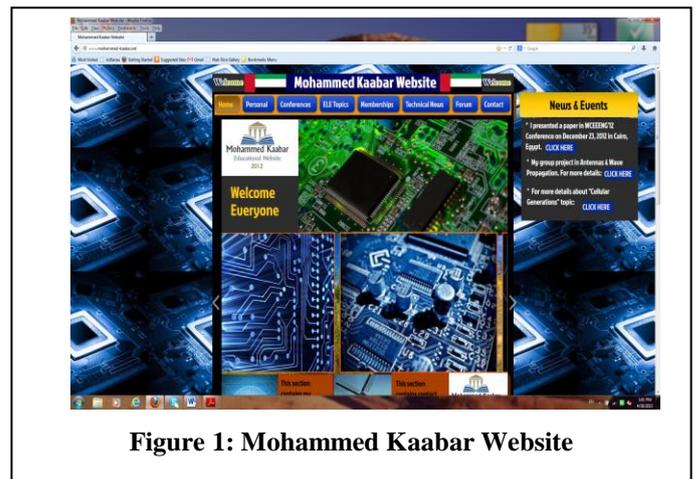

**Figure 1: Mohammed Kaabar Website**

The data was obtained from my own educational website called "Mohammed Kaabar Website" (http://www.mohammed-kaabar.net) by using the two HTML codes that can count the number of visitors from various countries worldwide. The data sets are available on a link on my website (http://http://www.mohammed-kaabar.net/#!contact/c2q4) .The first HTML code was created to only count the number of visitors as shown in Fig. 2.

```
<div align="center"><a
href="http://www.amazingcounters.com"><img border="0"
src="http://cc.amazingcounters.com/counter.php?i=3098543&c
=9295942" alt="web page visitor stats"></a><br><a
href="http://www.coupons-coupon-codes.com/lane-
bryant/">Lane Bryant Catalog</a></div>
```

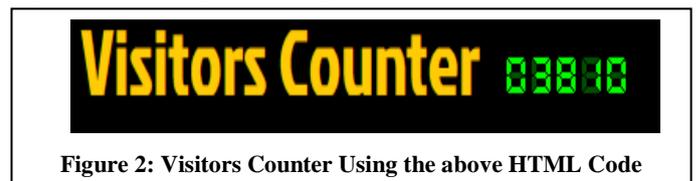

**Figure 2: Visitors Counter Using the above HTML Code**

Then, the second HTML code was created to count also number of visitors but it can show various properties of website visitors such as the types of browsers and operating systems used in visiting the website as shown in the next page.

```
<!-- Start of StatCounter Code for WebStarts -->
<script type="text/javascript">
var sc_project=8421830;
var sc_invisible=0;
var sc_security="1673b380";
var scJsHost = (("https:" == document.location.protocol) ?
"https://secure." : "http://www.");
document.write("<sc"+"ript type='text/javascript' src='" +
scJsHost +
"statcounter.com/counter/counter.js'></"+"script>");</script>
<noscript><div class="statcounter"><a title="web statistics"
href="http://statcounter.com/free-web-stats/"
target="_blank"><img class="statcounter"
src="https://c.statcounter.com/8421830/0/1673b380/0/"
alt="web statistics"></a></div></noscript>
<!-- End of StatCounter Code for WebStarts -->
```

From the above code, several sets of data were extracted in comma-separated values (CSV) files to be analyzed and visualized. Then, these data sets in CSV files were entered into Gephi as undirected graphs where the following are the criteria for the inclusion of nodes and edges for each data set.

***1-Browsers Types:*** each type of browser used by a certain visitor represents a node, and the number and percentage of hits by different visitors with the same types of browsers represent edge.

***2-Visitors' Country:*** each country represents a node, and the number and percentage of hits by different visitors within the same country represent edge.

***3-Visitors' City:*** each city represents a node, and the number and percentage of hits by different visitors within the same city represent edge.

***4-Visitors' State/Region:*** each state/region represents a node, and the number and percentage of hits by different visitors within the same state/region represent edge.

***5-The Combination of Visitors' Features Data:*** each node represents a visitor with all above types of data including the IP-addresses for that visitor, and edge represents the number and percentage of hits by different visitors with the same types of data.

### III. DATA ANALYSIS

All the above data sets will be analyzed and visualized using several metrics from SNA course in this paper.

### A- Browsers Types Data Analysis

The following table 1.1 shows some of these metrics for the types of browsers data and random graph as well.

| Metrics | Browser Types | Random Graph |
|---|---|---|
| Number of Nodes | 31 | 31 |
| Number of Edges | 30 | 30 |
| Number of Shortest Paths | 108 | 1038 |
| Average Path Length | 1.444 | 1.169 |
| Number of Communities | 8 | 9 |
| Diameter | 2 | 2 |
| Modularity | 0.791 | 0.133 |
| Average Weighted Degree | 1.935 | 13.903 |
| Mutual Degree Range | 1-5 | 1-5 |

**Table 1.1: Comparison between Browsers Types Graph and Random Graph**

According to the above table 1.1 which shows the comparison between both browsers types data graph and the random graph with the same number of nodes and edges where the number of shortest paths, number of communities and average weighed degree for random graph are greater than the Browsers Types graph with the same diameter and mutual degree range. However, the modularity and average path length of browsers types data is greater than the random graph because the division of a network into communities in these data is more strength with a dense connection between the nodes within the communities of browsers data types such as Safari, Firefox, Chrome and Android. Moreover, the Firefox has a higher modularity than the other communities as shown in the Fig. 3 below because a huge number of website visitors use this browser than other types of browsers.

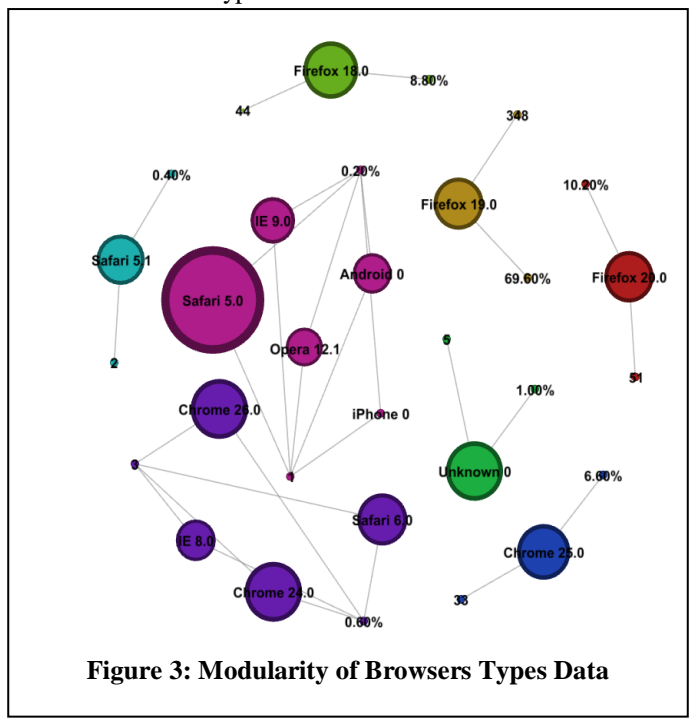

**Figure 3: Modularity of Browsers Types Data**

However, the modularity of the random graph with the same number of nodes and edges is also shown below in Fig. 4.

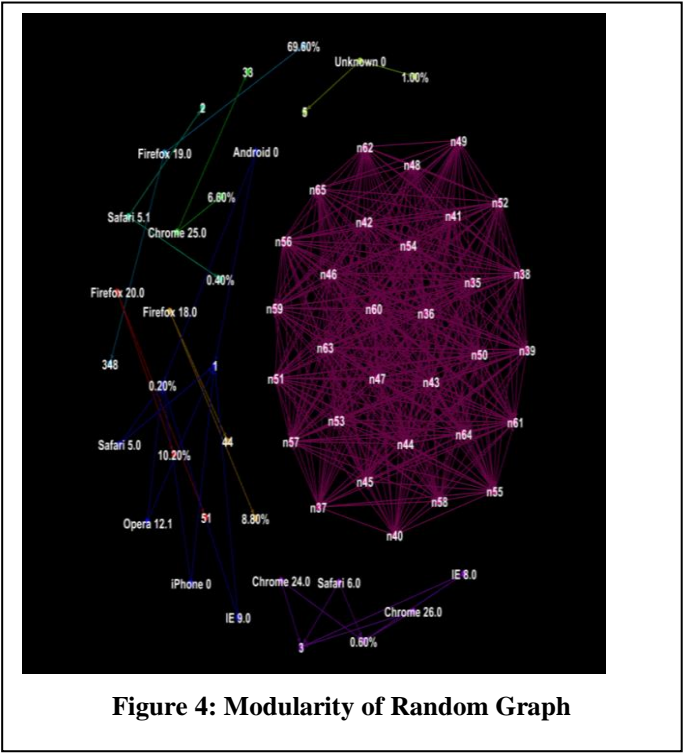

**Figure 4: Modularity of Random Graph**

In addition, as mentioned before, using the second HTML code, we can notice in Fig. 5 that the percentage of visitors who used the Firefox is 88.6% with 443 hits which is greater than all other browsers used by website visitors.

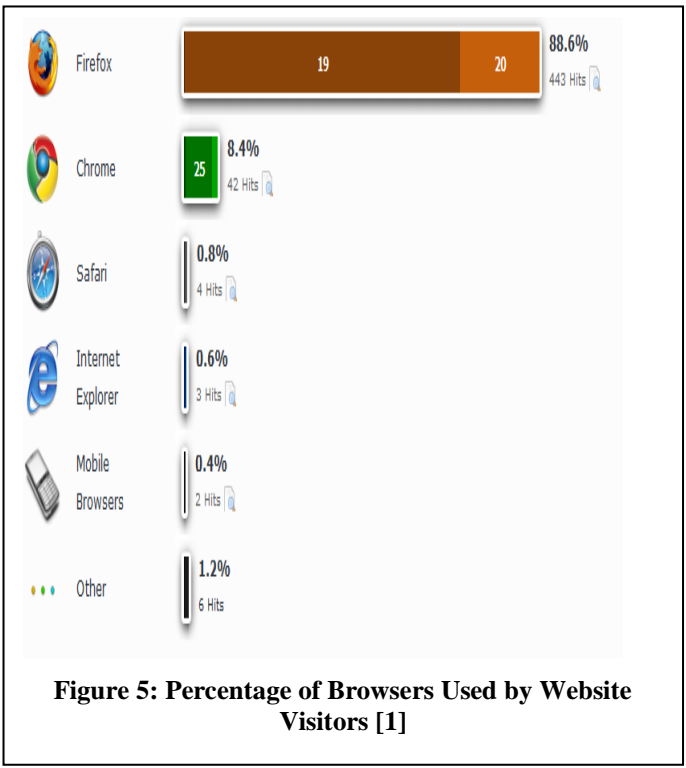

**Figure 5: Percentage of Browsers Used by Website Visitors [1]**

## B- Visitors' Country Data Analysis

The following table 2.2 shows some of these metrics for the visitors' country data and random graph as well.

| Metrics | Visitors' Country | Random Graph |
| --- | --- | --- |
| Number of Nodes | 35 | 35 |
| Number of Edges | 29 | 29 |
| Number of Shortest Paths | 350 | 1540 |
| Average Path Length | 1.834 | 1.316 |
| Number of Communities | 6 | 7 |
| Diameter | 2 | 2 |
| Modularity | 0.491 | 0.159 |
| Average Weighted Degree | 2.629 | 15.514 |
| Mutual Degree Range | 1-2 | 1-2 |

**Table 2.2: Comparison between Visitors' Country Graph and Random Graph**

According to the above table 2.2 which shows the comparison between both visitors' country graph and the random graph with the same number of nodes and edges where the number of shortest paths, number of communities and average weighed degree for random graph are greater than the visitors' country data graph with the same diameter and mutual degree range. However, the modularity and average path length of visitors' country data is greater than the random graph because the division of a network into communities in these data is more strength with a dense connection between the nodes within the communities of visitors' country. Furthermore, the United Arab Emirates has a higher modularity than the other communities as shown in the Fig. 6 below because a huge number of website visitors visit Mohammed Kaabar Website from this country than other countries.

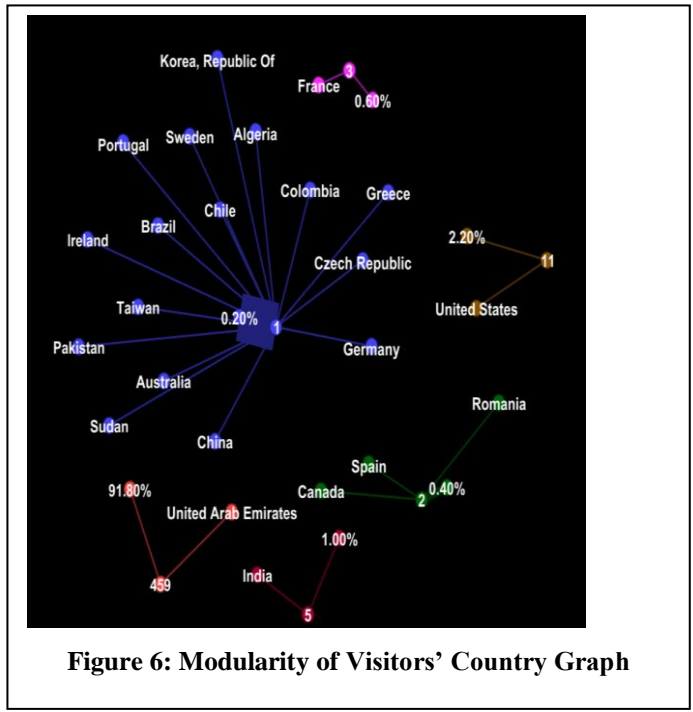

**Figure 6: Modularity of Visitors' Country Graph**

On the other hand, the visualization of the modularity of random graph with the same number of nodes and edges is also shown below in Fig. 7.

**Figure 7: Modularity of Random Graph**

Moreover, we can notice in Fig. 8 that the percentage of visitors who came from United Arab Emirates is 92.60% with 463 hits which is higher than other visitors coming from other countries worldwide.

**Figure 8: Modularity of Random Graph [1]**

## C- Visitors' City Data Analysis

The following table 3.3 shows some of these metrics for the visitors' city data and random graph as well.

| Metrics | Visitors' City | Random Graph |
|---|---|---|
| Number of Nodes | 99 | 99 |
| Number of Edges | 99 | 99 |
| Number of Shortest Paths | 9702 | 19404 |
| Average Path Length | 2.836 | 1.991 |
| Number of Communities | 3 | 3 |
| Diameter | 8 | 8 |
| Modularity | 0.266 | 0.073 |
| Average Weighted Degree | 3.071 | 43.404 |
| Mutual Degree Range | 1-77 | 1-77 |

**Table 3.3: Comparison between Visitors' City Graph and Random Graph**

According to the above table 3.3 which shows the comparison between both visitors' city graph and the random graph with the same number of nodes and edges where the number of shortest paths and average weighed degree for random graph are greater than the visitors' city data graph with the same diameter, number of communities and mutual degree range. However, the modularity and average path length of visitors' city data is higher than the random graph because the division of a network into communities in these data is more strength with a dense connection between the nodes within the communities of visitors' city. In addition, the Sharjah City in the United Arab Emirates has a higher modularity than the other communities as shown in the Fig. 9 below because a huge number of website visitors visit Mohammed Kaabar Website from this city than other cities worldwide.

**Figure 9: Modularity of Visitors' City Graph**

On the contrary, the visualization of the modularity of random graph with the same number of nodes and edges is also shown below in Fig. 10.

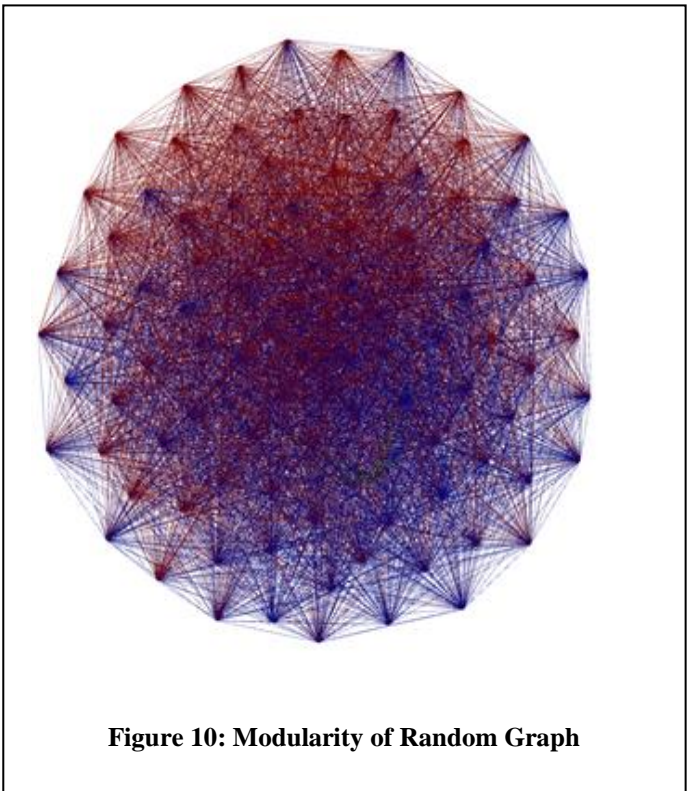

**Figure 10: Modularity of Random Graph**

Moreover, we can also notice in Fig. 11 below that the percentage of visitors who came from Sharjah City in the United Arab Emirates is 46.00% with 230 hits which is greater than other visitors coming from other cities worldwide.

**Figure 11: Modularity of Random Graph [1]**

## D- Visitors' State/Region Data Analysis

The following table 4.4 shows some of these metrics for the visitors' state/region data and random graph as well.

| Metrics | Visitors' State/Region | Random Graph |
|---|---|---|
| Number of Nodes | 70 | 70 |
| Number of Edges | 69 | 69 |
| Number of Shortest Paths | 4830 | 9660 |
| Average Path Length | 3.084 | 2.119 |
| Number of Communities | 3 | 4 |
| Diameter | 8 | 8 |
| Modularity | 0.269 | 0.101 |
| Average Weighted Degree | 3.229 | 30.814 |
| Mutual Degree Range | 1-51 | 1-51 |

**Table 4.4: Comparison between Visitors' State/Region Graph and Random Graph**

According to the above table 4.4 which shows the comparison between both visitors' state/region graph and the random graph with the same number of nodes and edges where the number of shortest paths, number of communities and average weighed degree for random graph are greater than the visitors' state/region data graph with the same diameter and mutual degree range. However, the modularity and average path length of visitors' country data is higher than the random graph because the division of a network into communities in these data is more strength with a dense connection between the nodes within the communities of visitors' state/region. Moreover, Sharjah has a higher modularity than the other communities as shown in the Fig. 12 below because a huge number of website visitors visit Mohammed Kaabar Website from this region than other regions/states.

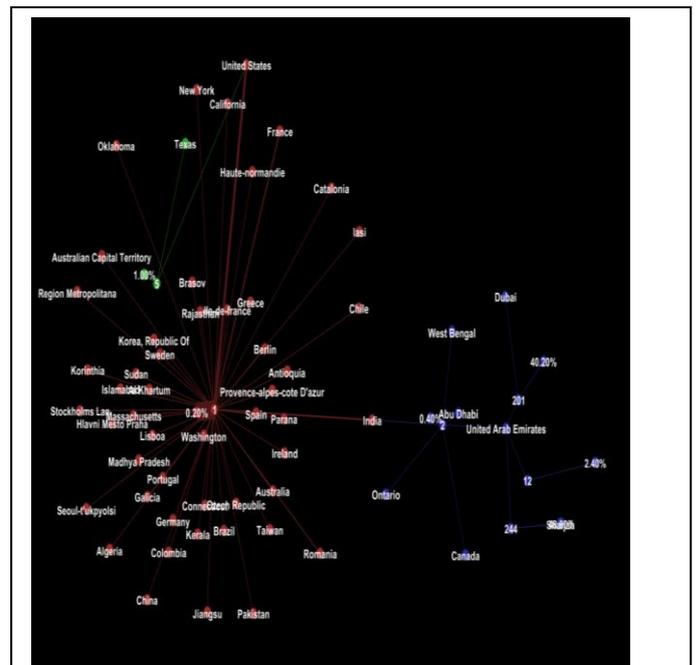

**Figure 12: Modularity of Visitors' State/Region Graph**

On the other hand, the visualization of the modularity of random graph with the same number of nodes and edges is also shown below in Fig. 13.

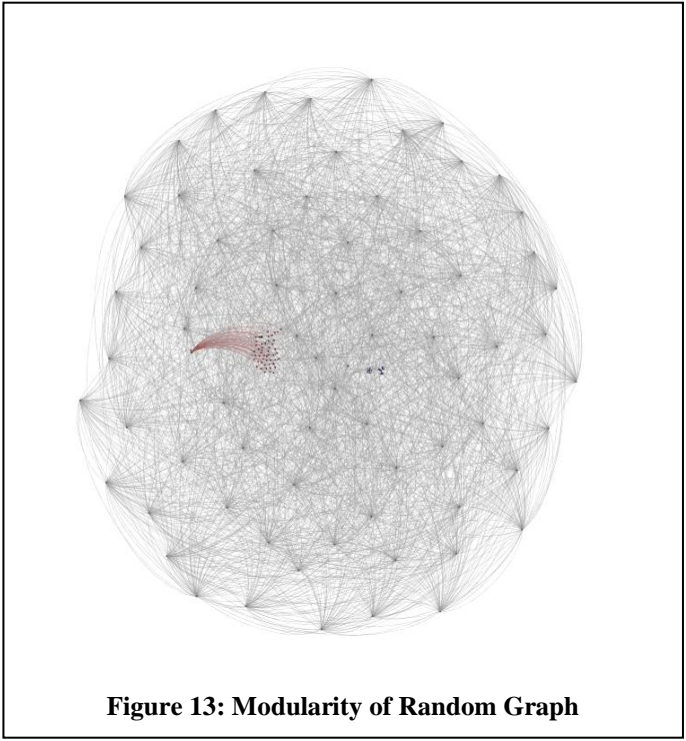

Figure 13: Modularity of Random Graph

### E- The Combination of Visitors' Features Data Analysis

The following table 5.5 shows some of these metrics for the combination of visitors' features data and random graph as well.

| Metrics | Visitors' Features | Random Graph |
|---|---|---|
| Number of Nodes | 1140 | 1140 |
| Number of Edges | 5514 | 5514 |
| Number of Shortest Paths | 1298460 | 2596920 |
| Average Path Length | 3.086 | 2.546 |
| Number of Communities | 4 | 3 |
| Diameter | 4 | 4 |
| Modularity | 0.265 | 0.265 |
| Average Weighted Degree | 10.488 | 33.685 |
| Mutual Degree Range | 1-500 | 1-500 |

Table 5.5: Comparison between Visitors' Features Graph and Random Graph

According to the above table 5.5 which shows the comparison between both visitors' features graph and the random graph with the same number of nodes and edges where the number of shortest paths, modularity and average weighed degree for random graph are higher than the visitors' features data graph with the same diameter and mutual degree range but the only difference among all above discussed data analysis is the number of communities for random graph is less than that for visitors' features. However, the average path length of visitors' features data is greater than the random graph because the division of a network into communities in these data is more strength with a dense connection between the nodes within the communities of visitors' features. Moreover, Fig. 14 below shows the modularity of visitors' features.

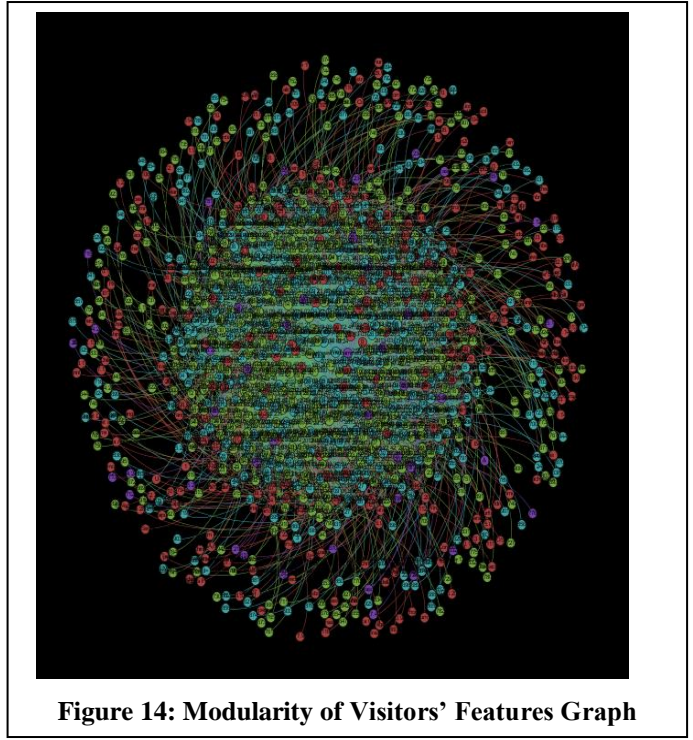

Figure 14: Modularity of Visitors' Features Graph

On the contrary, the visualization of the modularity of random graph with the same number of nodes and edges is also shown below in Fig. 15.

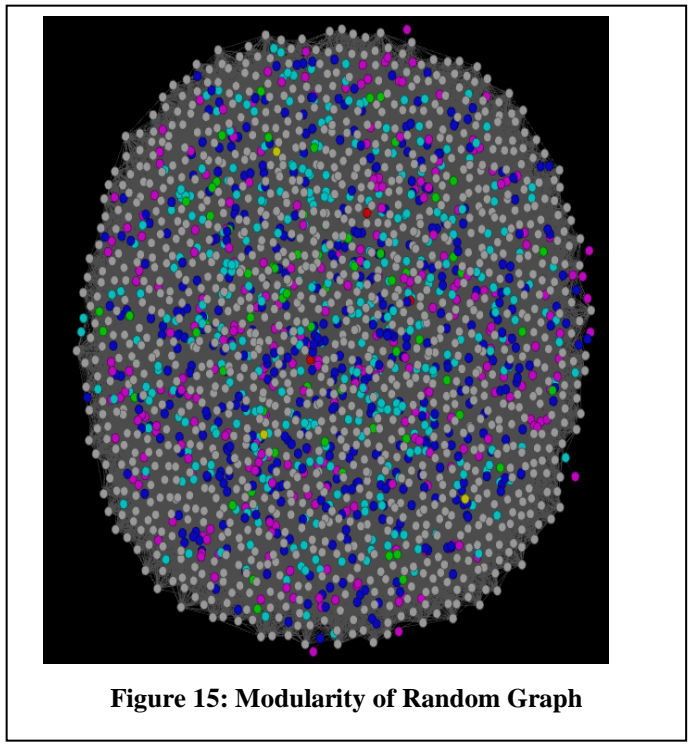

Figure 15: Modularity of Random Graph

## IV. INTREPRETATION

All the above data gave us a good understanding of how the empirical network analysis can be used for analyzing the website visitors through the methods used in the social network analysis. There are some limitations of the above analyzed data. First of all, one of these limitations is that the data of this project were collected with the number of visitors about 3810 which is not enough for a huge empirical network analysis. Second, data about the number of visitors who sign up for the forum in Mohammed Kaabar Website were not included in this project because the number of website visitors who signed up in forum was very small. Moreover, the data used in this project yield new insights where these analyzed data have potential to be used in the development of website by knowing which visitor visited this website several times and from which country, city, state/region, IP-address visited it as well as the spent period on website can also be determined using the above data. Therefore, my future work is to collect data with a large number of website visitors such as 10,000 visitors, and I will also include data about forum such as number of users who participates in forum in order to give an outstanding interpretation of empirical network analysis of website visitors.

## V. CONCLUSSION

All in all, this paper proves that website visitors' data can be analyzed as empirical network. In this project, several methods was used in order to analyze these data such as modularity, number of shortest paths, average path length, average weighed degree, number of communities, diameter, mutual degree range and number of nodes and edges. In addition, several visualizations about website visitors' data and random graph by Gephi were included in this project in order to compare both of these visualizations. As a result, these methods, visualizations, comparisons give a good understanding of how these data can be empirically analyzed and used for the development of website.

### ACKNOWLEDGMENT

I would like to thank Dr.Lada A. Adamic, Associate professor at the School of Information & Center for the Study of Complex Systems at the University of Michigan, for all her efforts for giving us a great opportunity to learn the methods used in social network analysis at the one of the top world universities (University of Michigan), and also for developing our skills in social network analysis (SNA) by giving us several examples, quizzes, PageRank competition and programming assignments where these activities make us experts in this field of study (Social Network Analysis) in order to do more research in this area. Ultimately, I would like to thank also all course stuff and course community for discussing great topics through forum threads.

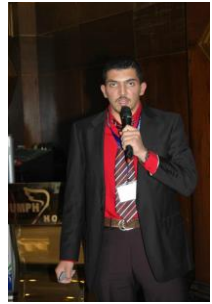

**Mohammed K.A Kaabar** is studying Bachelor of Science in the Electronics and Electrical Engineering. He became IEEE Student Member, IEEE Computer Society Member, IEEE Women Engineering Society Member and IEEE Communications Society Member, in 2011 and 2012, respectively. In 2011 and 2012, he presented a paper in the area of modern electronics in WCEEENG'12 conference in Cairo, Egypt. In 2013, he became a reviewer in the committee at the 8[th] Jordanian International Electrical and Electronics Engineering Conference (JIEEEC 2013). He also participated in several competitions such as writing a technical article on the environmental engineering technology, aiming at reducing the impacts of global warming. He also carried out several research papers and projects under the supervision of faculty titled "Louisiana in the Danger of Extinction: Analysis of the Problem of Global Warming to Find the Best Solution", "The Solar Energy in Our Future World: The Applications of the Solar Energy in Masdar City" and "The Benefits of Three Dimensional Transistor in the Microelectronics Area". In 2011, he attended a three-month course in numerical approximation techniques including error analysis, root finding, interpolation, function approximation, numerical differentiation, numerical integration and numerical solutions of initial value problems. Ultimately, he worked on several projects such as "PCA Implementation and Classification of Data in Recognition of Arabic Sign Language Alphabet using Polynomial Classifiers" and "Modeling a GaAs MESFET Device Structure using Silvaco Software: Athena and Atlas". For Further information about him, please .visit his persomal website: http:// www.mohammed-kaabar.net.